\documentclass[pdftex,twocolumn,epjc3]{svjour3}

\usepackage[utf8]{inputenc}
\usepackage{amsmath}
\usepackage{amssymb}
\usepackage[numbers,sort&compress]{natbib}
\usepackage{graphicx}
\usepackage[export]{adjustbox}
\usepackage[usenames,dvipsnames,svgnames,table]{xcolor}
\usepackage{multirow}

\usepackage[colorlinks,citecolor=blue,urlcolor=blue,linkcolor=blue]{hyperref}

\newcommand{\eref}[1]{Eq. (\ref{#1})}
\newcommand{\fref}[1]{Fig. \ref{#1}}
\newcommand{\tref}[1]{Tab.~\ref{#1}}
\newcommand{\nnnl}{\nonumber\\}	

\allowdisplaybreaks 

\journalname{Eur. Phys. J. C}

\begin{document} 

\title{Spectrum of scalar and pseudoscalar glueballs from functional methods}

\author{Markus Q.~Huber\thanksref{e1,addr1} \and Christian S. Fischer\thanksref{e2,addr1,addr2} \and H\`elios Sanchis-Alepuz\thanksref{e3,addr3,addr4}}

\thankstext{e1}{e-mail: markus.huber@physik.jlug.de}
\thankstext{e2}{e-mail: christian.fischer@theo.physik.uni-giessen.de}
\thankstext{e3}{e-mail: helios.sanchis-alepuz@silicon-austria.com}

\institute{Institut f\"ur Theoretische Physik, Justus-Liebig-Universit\"at Giessen, 35392 Giessen, Germany\label{addr1}
          \and
          Helmholtz Forschungsakademie Hessen f\"ur FAIR (HFHF),
	      GSI Helmholtzzentrum f\"ur Schwerionenforschung, Campus Gie{\ss}en, 35392 Gie{\ss}en, Germany\label{addr2}
          \and
          Institute of Physics, University of Graz, NAWI Graz, Universit\"atsplatz 5, 8010 Graz, Austria\label{addr3}
          \and
          Silicon Austria Labs GmbH, Inffeldgasse 33, 8010 Graz, Austria\label{addr4}
}

\date{\today}

\maketitle

\begin{abstract}
We provide results for the spectrum of scalar and pseudoscalar glueballs in pure Yang-Mills theory using a
parameter-free fully self-contained truncation of Dyson-Schwinger and Bethe-Salpeter equations. The only input, 
the scale, is fixed by comparison with lattice calculations. We obtain 
ground state masses of $1.9\,\text{GeV}$ and $2.6\,\text{GeV}$ for the scalar and pseudoscalar glueballs, 
respectively, and $2.6\,\text{GeV}$ and $3.9\,\text{GeV}$ for the corresponding first excited states. This is 
in very good quantitative agreement with available lattice results. Furthermore, we predict masses for the second 
excited states at $3.7\,\text{GeV}$ and $4.3\,\text{GeV}$.
The quality of the results hinges crucially on the self-consistency of the employed input.
The masses are independent of a specific choice for the infrared behavior of the ghost propagator providing further evidence that this only reflects a nonperturbative gauge completion.
\end{abstract}

\section{Introduction}

Glueballs, i.e. hadrons that consist of gluons only, are extremely fascinating objects to study. They arise 
due to the non-Abelian nature of Yang-Mills theory which allows for the formation of gauge invariant states of 
gluons that interact strongly amongst each other. The properties of glueballs have been studied in many
models since their prediction in the 1970s \cite{Fritzsch:1972jv,Fritzsch:1975tx}. Today, many glueball masses 
in pure Yang-Mills theory are known rather accurately owing to high statistics quen\-ched lattice calculations
\cite{Bali:1993fb,Morningstar:1999rf,Chen:2005mg,Athenodorou:2020ani}. Unquenched lattice calculations of glueball masses are still 
on the exploratory level with considerable uncertainties due to severe problems with the signal to noise ratio,
see, e.g.
\cite{Gregory:2012hu} and references therein.
Alternative theoretical frameworks, such as Hamiltonian many body methods \cite{Szczepaniak:1995cw,Szczepaniak:2003mr} 
or chiral Lagrangians \cite{Janowski:2011gt,Eshraim:2012jv}, have shed some light on potential mass patterns and 
identifications of experimental states dominated by their glueball content, see \cite{Ochs:2013gi} for a comprehensive 
review. However, it seems fair to state that our detailed understanding of glueball formation from the underlying 
dynamics of Yang-Mills theory is still far from complete. 
In this work, we provide an additional, complementary perspective from functional methods. 

While the calculation of mesons and baryons from functional bound state equations is a very active field, see, e.g.
\cite{Cloet:2013jya,Eichmann:2016yit} and references therein, studies of exotic states in this framework are less 
abundant. This is particularly true for glueballs due to the inherent complexity of gauge-fixed Yang-Mills theories.
Some results have been reported in \cite{Meyers:2012ka,Sanchis-Alepuz:2015hma,Souza:2019ylx,Kaptari:2020qlt}, but they remain on an exploratory level due to the ansaetze used for the input.
An alternative approach to extract glueball masses from Landau gauge correlation functions was followed in \cite{Dudal:2010cd,Dudal:2013wja}.

\begin{figure*}[tb]
	\includegraphics[width=0.98\textwidth]{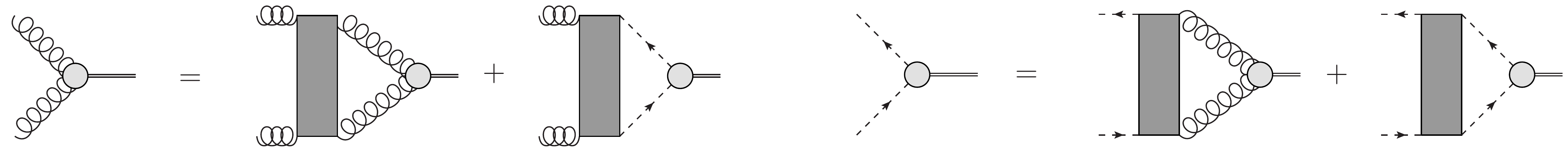}
	\caption{
		The coupled set of BSEs for a glueball made from two gluons and a pair of Faddeev-Popov (anti-)ghosts.
		Wiggly lines denote dressed gluon propagators, dashed lines denote dressed ghost propagators. 
		The gray boxes represent interaction kernels given in Fig.~\ref{fig:kernels}. The Bethe-Salpeter amplitudes of the glueball
		are denoted by gray disks.
	}
	\label{fig:bses}
\end{figure*}

Functional methods allow direct studies of the internal structures of bound states as determined by the dynamics of QCD.
As an example, let us mention tetraquarks where the role of different two-body clusterings can be studied, see Ref.~\cite{Eichmann:2020oqt} for a summary.
Another promising aspect of functional methods is that they provide direct access to analytic properties.
The necessary numerical techniques are not fully developed yet, but they have already been applied to some interesting questions like dynamical resonances or spectral functions, see, for instance, \cite{Maris:1995ns,Alkofer:2003jj,Windisch:2012sz,Strauss:2012dg,Windisch:2013dxa,Pawlowski:2015mia,Strodthoff:2016pxx,Eichmann:2019dts,Fischer:2020xnb}.
As a third example where functional methods can provide useful insights into QCD we mention the study of its phases and the transitions between them, see, e.g., \cite{Pawlowski:2010ht,Fischer:2018sdj}.
In the present work we add another application of functional equations to the list.
It is set apart from the previous examples by delivering quantitative results without the need for any model parameters which typically substitute for missing information in the considered truncated systems of equations.
This is to our knowledge the first such calculation.
If such setups can be devised and solved also for other quantities of interest, this would constitute another major step forward.

What enabled us to perform a parameter-free calculation is the solid understanding of the properties of Yang-Mills correlation functions obtained in the last decade and a refinement of the methods to compute them from functional equations, e.g. 
\cite{Huber:2012kd,Blum:2014gna,Eichmann:2014xya,Huber:2016tvc,Cyrol:2016tym,Huber:2017txg,Aguilar:2018csq,Huber:2018ned,Aguilar:2019jsj,Aguilar:2019kxz,Huber:2020keu}.
In particular, we employ recent results from a fully self-contained truncation of Dyson-Schwinger equations (DSEs) for
propagators and vertices that leads to good agreement with corresponding gauge fixed lattice results without 
any tuning \cite{Huber:2020keu}. As we will see in the following, the glueball spectrum extracted from the 
corresponding set of bound state Bethe-Salpeter equations (BSEs) agrees quantitatively with corresponding lattice results.

The remainder of the article is organized as follows.
In Sec.~\ref{sec:framework} we introduce the bound state equations we solve and the employed input.
The extraction of the spectrum is explained in Sec.~\ref{sec:spectrum_extraction} and the results are presented and discussed in Sec.~\ref{sec:results}.
We summarize and provide an outlook in Sec.~\ref{sec:summary}.
The employed methods are illustrated in the appendix for a meson system.

\section{Framework and input}
\label{sec:framework}


\begin{figure}[tb]
 \includegraphics[width=0.48\textwidth]{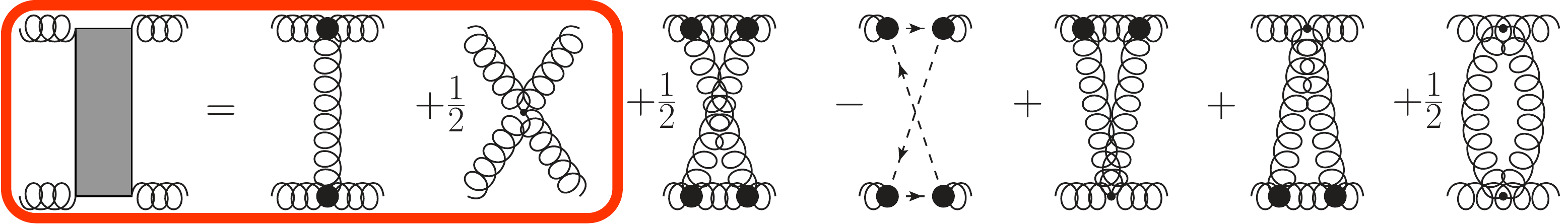}\\ 
 \vskip4mm
 \includegraphics[height=1.2cm]{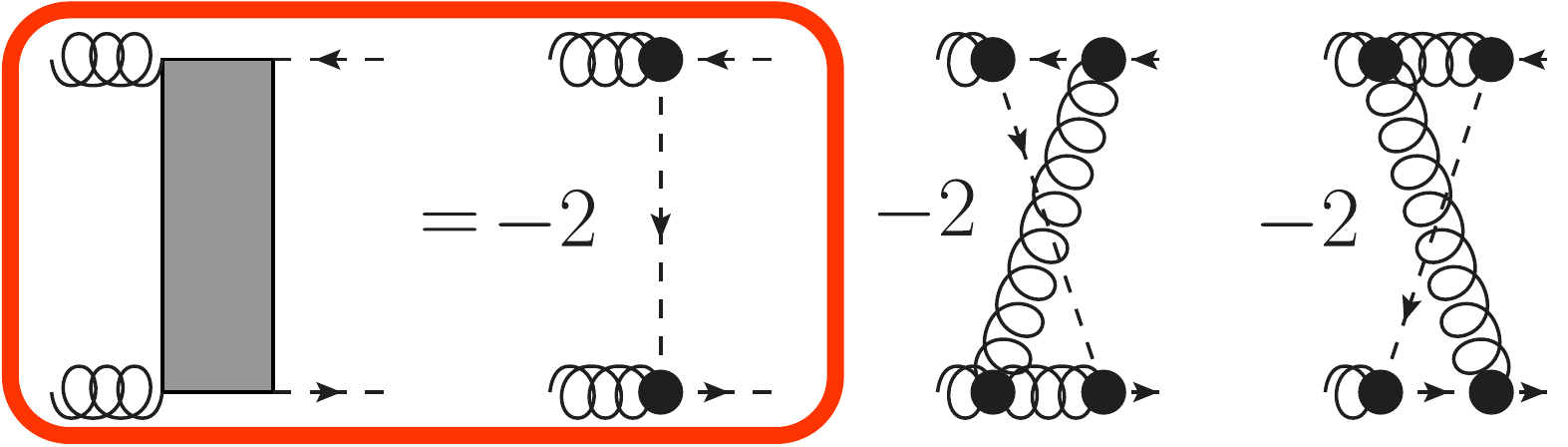}\hfill
 \includegraphics[height=1.2cm]{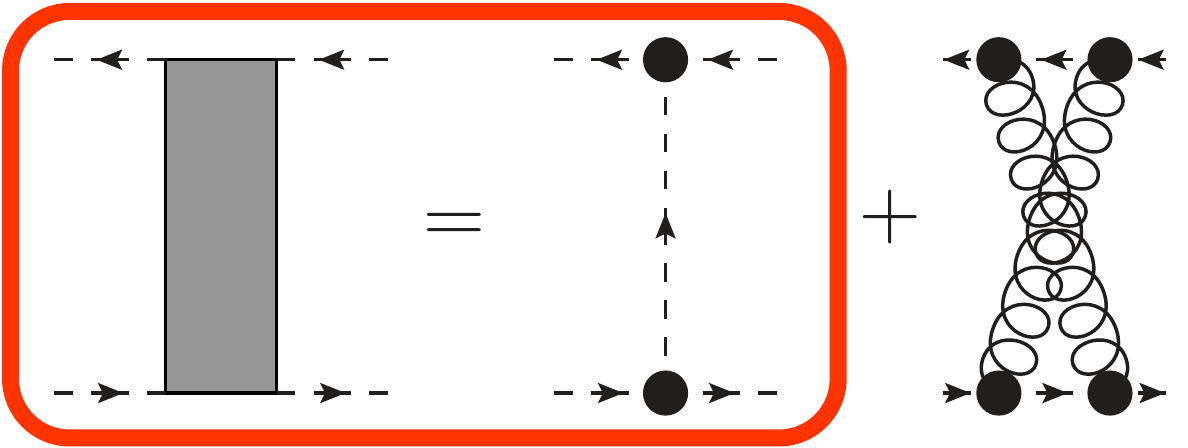}\\
 \vskip4mm
 \includegraphics[height=1.2cm]{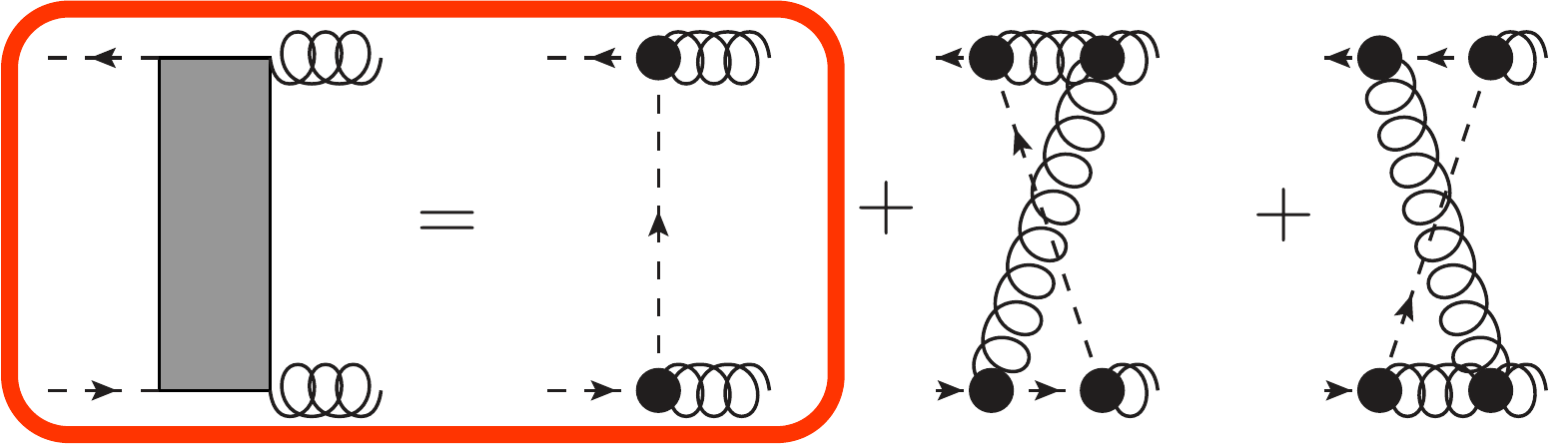}
 \caption{
 Interaction kernels from the three-loop 3PI effective action.
 All propagators are dressed; black disks represent dressed vertices.
 In our calculation, we include the diagrams inside the red rectangles. 
 Details are discussed in the text.}
 \label{fig:kernels}
\end{figure}

We consider Landau gauge Yang-Mills theory with gauge fixing via the Faddeev-Popov procedure. The corresponding bound 
state equation for a two-gluon glueball is depicted in \fref{fig:bses}. Since for some quantum numbers 
the two-gluon state also couples to a ghost--anti-ghost state, we also need to consider the corresponding BSE.
For simplicity, we refer to them as glueball-part and ghostball-part.
The forms of the interaction kernels can be derived from an effective action and its truncation.
We use the kernels obtained from a 3PI effective action truncated to three loops \cite{Berges:2004pu,Carrington:2010qq}.
This choice is motivated by excellent results that have been obtained in this truncation scheme for the coupled set
of DSEs for the dressed gluon propagator, the dressed ghost propagator and the dressed ghost-gluon as well as 
three-gluon vertices \cite{Huber:2020keu} discussed below. A truncation on a similar level has also been used in the
quark sector in Ref.~\cite{Williams:2015cvx}. The corresponding kernels, shown in \fref{fig:kernels}, are derived by 
performing appropriate functional derivatives of the action \cite{Fukuda:1987su,McKay:1989rk,Sanchis-Alepuz:2015hma,Sanchis-Alepuz:2015tha}.

There are some noteworthy and important differences between a setup based on the 1PI effective action and our 3PI setup.
The 3PI effective action depends on dressed propagators and three-point functions which can be calculated from their equations of motion corresponding to stationarity conditions of the action.
Typically, one employs a (dressed) loop expansion for the 3PI effective action with a finite number of terms \cite{Berges:2004pu,Carrington:2010qq}. The truncated DSE system derived therefrom is then a finite and closed system of equations for propagators and three-point functions with all higher-order vertices bare. Thus, it contains non-perturbative information for the three-point functions in a significantly different fashion than a truncation derived from the 1PI effective action. In the latter, all propagators and vertices are treated on the same footing, and for any truncation of the 1PI effective action, there are infinitely many equations to be solved.
Thus, one normally performs the truncation at the level of the system of equations (derived from the 1PI effective action) instead of the action itself.
If the 3PI effective action is truncated to three loops, the equations of motions for the three-point functions have a similar structure as the equations from the 1PI effective action without two-loop terms \cite{Berges:2004pu}.
However, there are no bare three-point functions in the former case.
This seems like a small difference, but it turned out that the equations of motion of the three-loop expanded 3PI effective action, which are one-loop, outperform the results from equations of 1PI equations truncated to one loop.
An explicit comparison for the three-gluon vertex showed that including the two-loop diagrams for the latter leads to good agreement.
However, such two-loop calculations are much harder to realize \cite{Huber:2020keu}.

The similar structure of the equations of motion from 1PI and 3PI effective actions persists also for the kernels of the bound state equations.
A typical kernel for a 1PI calculation contains only one-particle exchanges with one bare vertex, but some studies beyond that exist as well, e.g., \cite{Bender:1996bb,Fischer:2009jm}.
The lowest order of 3PI kernels contain only one-particle exchanges but in contrast to the 1PI case all vertices are dressed.
The next order also contains two-particle exchanges.
For now, these are not taken into account due to their technical complexity. 

Of course, a priori there is no known way to guarantee that higher order corrections are small in a dressed loop expansion. 
The results of this work, in particular the excellent agreement of our glueball masses with corresponding lattice results, 
provides some circumstantial evidence that this may be the case. Further evidence has been collected in \cite{Eichmann:2014xya,Huber:2016tvc,Huber:2017txg,Huber:2018ned}, where selected extensions of the present truncation scheme
have already been calculated and found to be small indeed. Nevertheless, more systematic tests in dedicated calculations need 
to be done in the future to further elaborate on this issue.

Having specified the underlying equations, we can now turn to solving them.
A solution to the coupled set of BSEs for glueballs can be found by treating them as an eigenvalue equation for the matrix $\mathcal{K}$ of interaction kernels with the eigenvector $\Gamma=(\Gamma_{\mu\nu},\Gamma_{gh})$
combining the glueball-part $\Gamma_{\mu\nu}$ and the ghostball-part $\Gamma_{gh}$ of the Bethe-Salpeter amplitude. 
This leads to 
\begin{align}
\label{eq:eigenvalue_eq}
 \mathcal{K}\cdot\Gamma(P,p)=\lambda(P)\,\Gamma(P,p),
\end{align}
where $P$ and $p$ are the total and relative momenta.
For a solution, the eigenvalue $\lambda(P)$ must equal $1$.
The mass is determined from the corresponding value of the total momentum, $M^2=-P^2$.

This structure is very general and valid for all allowed quantum numbers.
Here we treat the simplest glueballs, the scalar and pseudoscalar ones.
To this end, we need to specify the forms for the amplitudes.
The glueball-part $\Gamma_{\mu\nu}$ has two open Lorentz indices and is transverse. For the scalar glueball ($J^{PC}=0^{++}$) one can find two independent tensors with these
properties given by \cite{Meyers:2012ka}:
\begin{align}\label{scalar1}
&\Gamma^{++}_{\mu\nu}(p,P)=h^{++}_2(p,P)\left(g_{\mu\nu}-\frac{p_{2\mu}p_{1\nu}}{p_1\cdot p_2}\right)\nnnl
&\quad +h^{++}_1(p,P)\frac{(p_1^2 \,p_{2\mu}-p_1\cdot p_2\,p_{1\mu})(p_2^2\, p_{2\nu}-p_{1}\cdot p_2\, p_{1\nu})}{(p_1\cdot p_2)^3-p_1\cdot p_2\, p_1^2\, p_2^2}
\end{align}
with $p_{1/2}=p\pm P/2$.
For the pseudoscalar glueball ($J^{PC}=0^{-+}$), only one tensor exists which we choose as
\begin{align}
 \Gamma^{-+}_{\mu\nu}(p,P)=h^{-+}(p,P)\,\epsilon_{\mu\nu\rho\sigma}\hat{p}^T_\rho \hat{P}_\sigma.
\end{align}
The hat indicates normalization and the superscript $T$ that the vector is made transverse with respect to $P$.
For the ghostball-part, which is a scalar in Lorentz space, the amplitude is simply given by
\begin{align}\label{scalar2}
 \Gamma^{++}_{gh}(p,P)=h^{++}_3(p,P).
\end{align}
There is no corresponding amplitude with negative parity.
This simplifies the BSE for the pseudoscalar glueball where the ghostball-part of the amplitude drops out.

The input required to solve the BSEs are the dressed gluon and ghost propagators, $D_{\mu\nu}$ and $D_G$, respectively, given by
\begin{equation}
D_{\mu \nu}(p) = \left(\delta_{\mu\nu}-\frac{p_\mu p_\nu}{p^2}\right) \frac{Z(p^2)}{p^2}, \,\,\,\,\,\,\,D_G(p)=\frac{-G(p^2)}{p^2},
\end{equation}
as well as the dressed three-gluon 
and ghost-gluon vertices. For these we use numerical results from a DSE system also derived from the three-loop 3PI 
effective action. A graphical representation of this truncation together with a thorough discussion of all technical 
details and merits can be found in \cite{Huber:2020keu}. Here, we only wish to mention that the scheme is self-contained,
i.e., it can be solved without any ad-hoc ansaetze and parameters. Thus, either correlation functions are taken into account 
and solved for self-consistently, or they are consistently neglected.

\begin{figure*}[t]
	\includegraphics[width=0.48\textwidth]{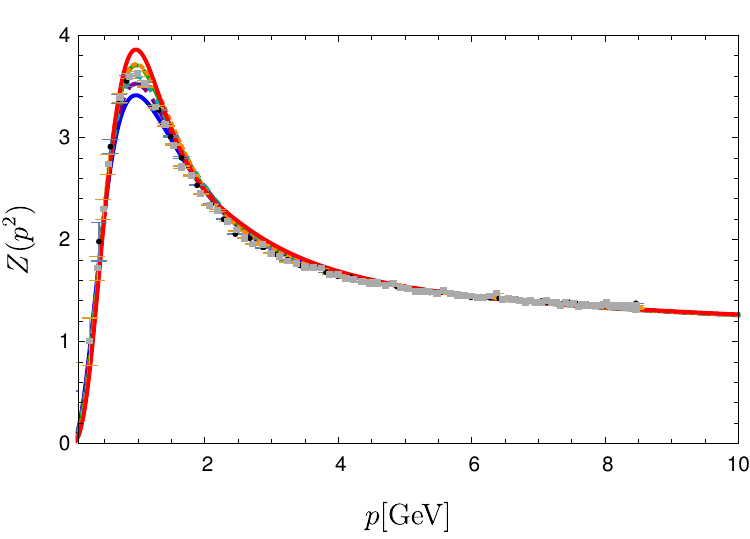}\hfill
	\includegraphics[width=0.48\textwidth]{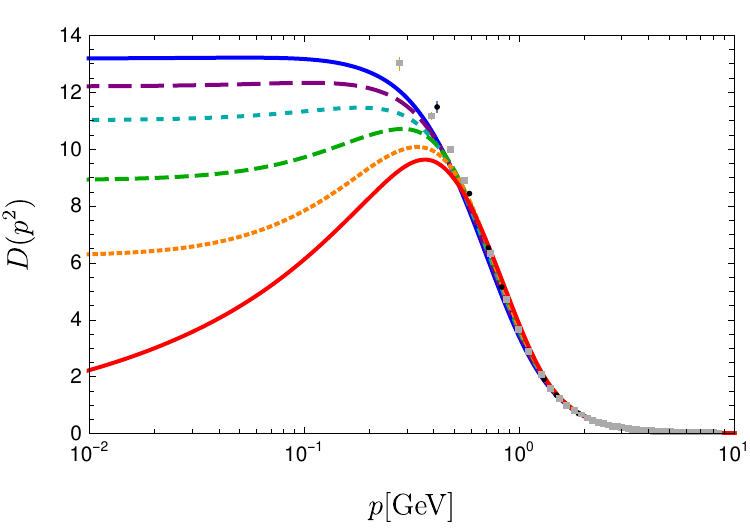}
	\caption{Gluon dressing function $Z(p^2)$ (left) and gluon propagator $D(p^2)$ (right) in comparison to lattice data \cite{Sternbeck:2006rd}.
	For the sake of comparison, the functional results were renormalized to agree with the lattice results at $6\,\text{GeV}$.
    Different lines correspond to different decoupling/scaling solutions as explained in the text.}
	\label{fig:gluon}
\end{figure*}

\begin{figure*}[t]
	\includegraphics[width=0.48\textwidth]{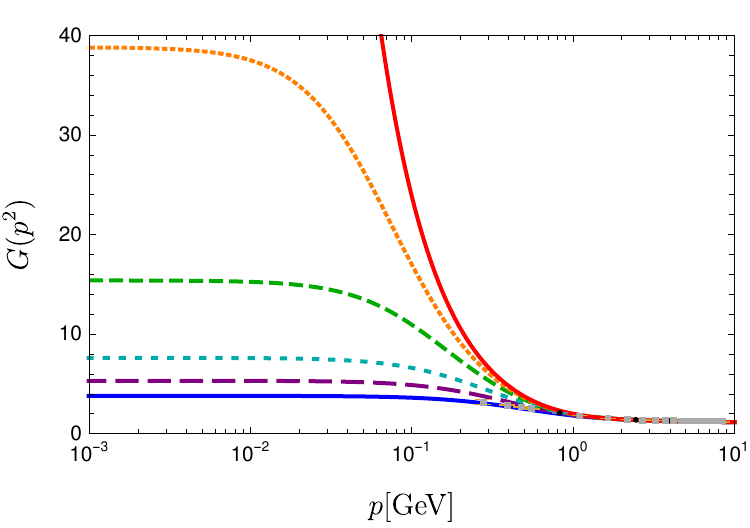}\hfill
	\includegraphics[width=0.48\textwidth]{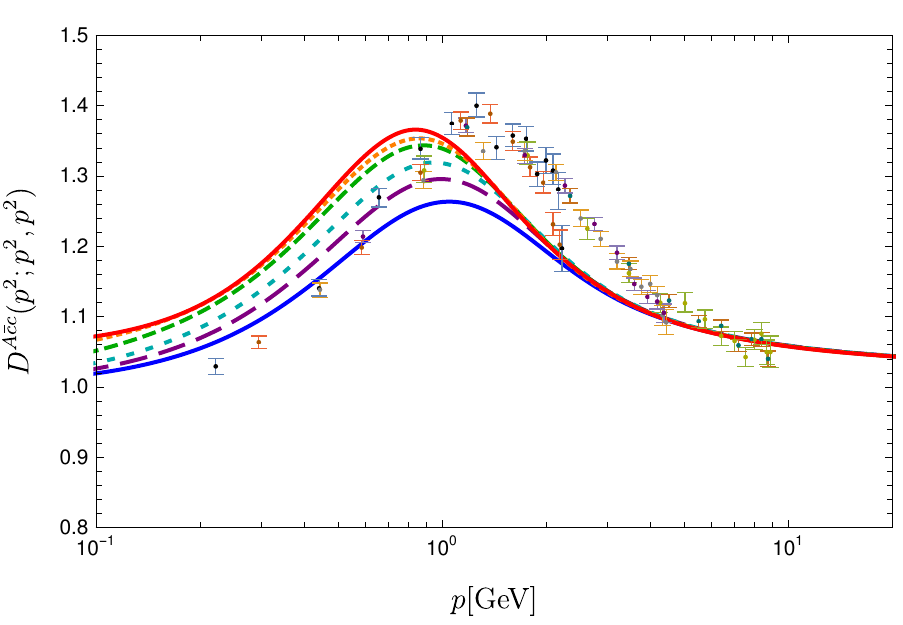}
	\caption{Left: Ghost dressing function $G(p^2)$ in comparison to lattice data \cite{Sternbeck:2006rd}.
	For the sake of comparison, the functional results were renormalized to agree with the lattice results at $8\,\text{GeV}$.
	Right: Ghost-gluon vertex dressing function at the symmetric point in comparison to $SU(2)$ lattice data \cite{Maas:2019ggf}.
    Different lines correspond to different decoupling/scaling solutions as explained in the text.}
	\label{fig:ghost}
\end{figure*}
    
\begin{figure}[t]
	\includegraphics[width=0.48\textwidth]{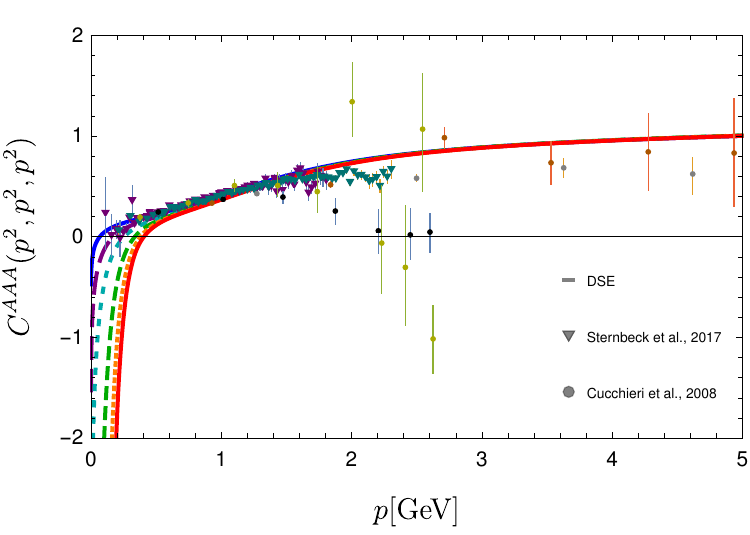}
	\caption{Solutions for the three-gluon vertex dressing function at the symmetric point in comparison to lattice data \cite{Cucchieri:2008qm,Sternbeck:2017ntv}).
	For the sake of comparison, all data were renormalized to $1$ at $5\,\text{GeV}$.
    Different lines correspond to different decoupling/scaling solutions as explained in the text.}
	\label{fig:3g}
\end{figure}

A truncation independent property of the Yang-Mills correlation functions in the continuum is that they appear as a 
one-parameter family of so-called decoup\-ling/scaling solutions with the scaling limit as an end-point
\cite{Boucaud:2008ji,Aguilar:2008xm,Fischer:2008uz,Alkofer:2008jy}. For the ghost and gluon propagators, these 
solutions agree at large momenta but start to differ around the scale of $1\,\text{GeV}$ and below. The appearance
of this one-parameter family has been discussed to be caused by incomplete nonperturbative gauge fixing in the Landau gauge/Faddeev-Popov setup
due the Gribov problem \cite{Fischer:2008uz,Maas:2009se}. Indeed, on the lattice it is well studied 
that different samplings of the Gribov region influence the low momentum behavior of correlation functions, see, e.g.
\cite{Maas:2009se,Maas:2011se,Sternbeck:2012mf} and references therein.
If this picture is correct, all solutions of the family should lead to the same results for physical 
quantities. The glueballs studied in this work provide an ideal testing ground for this hypothesis.

We thus used a number of different solutions within this one-parameter family including the scaling solution taken from Ref.~\cite{Huber:2020keu} for our calculation of glueball masses.
The corresponding gluon propagators and dressing functions are shown in Fig.~\ref{fig:gluon}.
Since the precise correspondence of continuum and lattice gauge fixing methods is an open issue \cite{Maas:2019ggf}, we show a selection of functional results and lattice results for one particular prescription to deal with Gribov copies.
All employed gluon propagators feature a maximum at $p^2>0$, but they are so shallow for two cases that they are hardly visible in the plots.
The ghost dressing function is shown in Fig.~\ref{fig:ghost}.
For the plots, the propagators were renormalized such that they agree with the lattice data of Ref.~\cite{Sternbeck:2006rd} at some high momentum.
This rescaling is unrelated to the renormalization employed in the actual calculation, where different renormalization conditions were used.
Especially for the vertices the consistency of the renormalization procedure is an important issue as discussed in \cite{Huber:2020keu}.
Here, we can directly use these renormalized results.

The ghost-gluon vertex dressing function is shown in Fig.~\ref{fig:ghost}.
The plot only shows the symmetric point at which all momentum squares are equal, but its full kinematic dependence was taken into account in the calculation.
The ghost-gluon vertex is finite in the Landau gauge \cite{Taylor:1971ff} and thus no renormalization is necessary.
The three-gluon vertex is shown in Fig.~\ref{fig:3g} where it is compared to lattice data from Refs.~\cite{Cucchieri:2008qm,Sternbeck:2017ntv}; see Refs.~\cite{Athenodorou:2016oyh,Boucaud:2017obn} for similar results.
Also here the full kinematic dependence was taken into account, but its angular dependence is very weak.
For the sake of comparison, all data were renormalized to $1$ at $5\,\text{GeV}$.

With the functional renormalization group, a truncation very similar to the one used here leads to equally good results
for the propagators and vertices \cite{Cyrol:2016tym} which further supports the trustworthiness of the present truncation.
While improvements of this truncation were partially already tested and found to be small, e.g., the impact of neglected
contributions like four-point functions \cite{Huber:2016tvc,Huber:2017txg,Huber:2018ned} and a full basis for the three-gluon
vertex \cite{Eichmann:2014xya}, more tests should be performed in the future.

For quark anti-quark bound states, the consistency of the truncations for the DSE and BSE system 
is known to be crucial. This is true in particular for the chiral properties, where the axial-vector Ward-Takahashi 
identity relates the self-energies and integration kernels, see, e.g. \cite{Eichmann:2016yit}. In the present case
of pure Yang-Mills glueballs, similar relations do not exist to our knowledge, but it is certainly important to consider 
the impact of consistency between the truncations of the DSEs and the BSEs.

\begin{figure*}[tb]
	\includegraphics[width=0.48\textwidth]{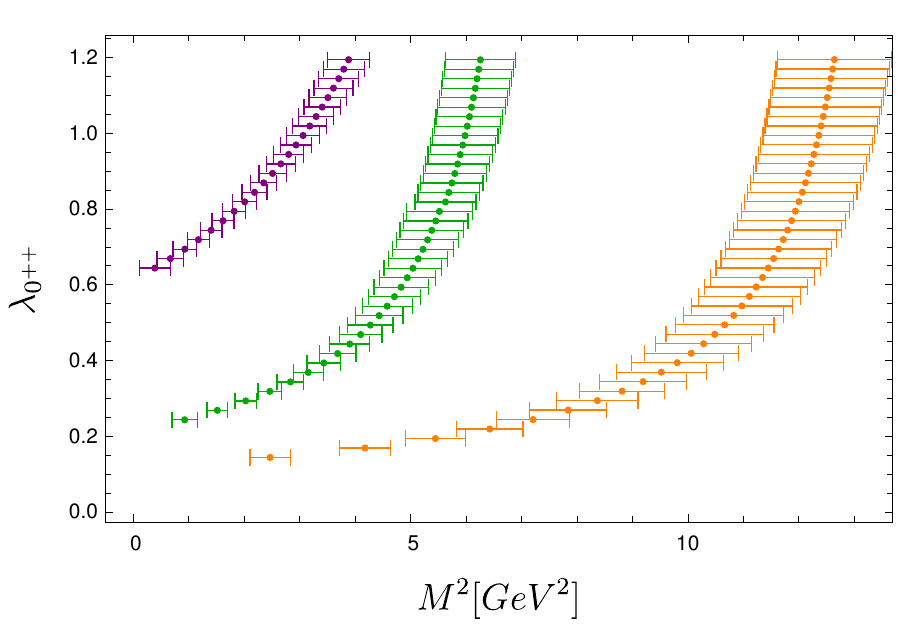}\hfill
	\includegraphics[width=0.48\textwidth]{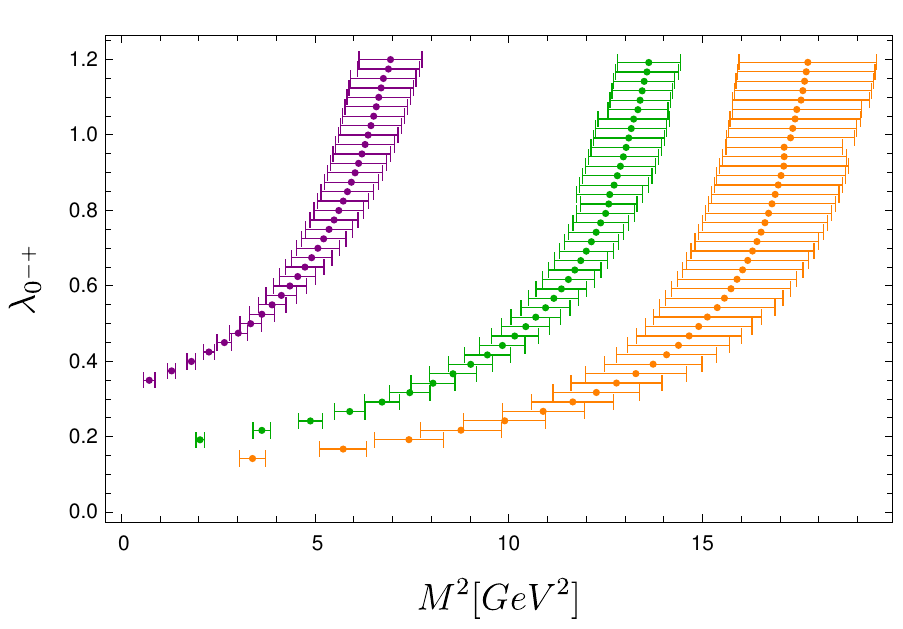}
	\caption{Eigenvalues for off-shell values of $P^2=-M^2$ for scalar (left) and pseudoscalar (right) glueballs.
		Shown are the eigenvalues for the ground state and the first two excited states.}
	\label{fig:extrapolation_eigenvalues} 
\end{figure*}

\section{Extraction of the spectrum}
\label{sec:spectrum_extraction}

Bethe-Salpeter equations describe bound states with time-like total momenta. As a consequence, propagators and vertices 
inside Bethe-Salpeter equations are tested at complex squared momenta. While calculations in the complex plane are in principle 
possible with functional methods, see, e.g. \cite{Maris:1995ns,Fischer:2005en,Fischer:2008sp,Krassnigg:2009gd,Strauss:2012dg,Kamikado:2013sia,Pawlowski:2015mia,Fischer:2020xnb}, such 
calculations are realized only with comparatively simple truncations. More advanced truncation schemes, like the one we use 
in this work, have been solved for space-like Euclidean momenta only. Analytic continuations of correlation functions have 
been explored \cite{Dudal:2013yva,Cyrol:2018xeq,Binosi:2019ecz,Dudal:2019gvn}, but definite conclusions are still lacking.   

Thus, we follow here an alternative approach and instead analytically continue the eigenvalues calculated for Euclidean momenta $P^2 >0$ into 
the time-like momentum domain $P^2<0$.
For every bound state we have calculated the eigenvalue $\lambda(P^2)$ of the BSE for $100$ space-like and real momenta $P^2 \in [10^{-4},0.25]\,\text{GeV}^2$
and extrapolate to time-like momenta using Schles\-sin\-ger's method based on continued fractions \cite{Schlessinger:1968spm,Tripolt:2018xeo}.
The extrapolation error is dominated by (small) numerical inaccuracies in the input data and the limitations of the chosen extrapolation 
function itself. We partially quantify the extrapolation error by a bootstrap-like procedure:
We take only a random subset of 80 points, which is enough to get a good extrapolation function, and calculate the mass. Excluding exceptional 
extrapolations that do not give a bound state, we repeat that process 100 times and average the results. The one standard deviation 
errors indicated in our results for the bound state masses stem from this procedure. Of course, this does not include additional 
errors due to the truncation and other sources. 

We tested the extrapolation procedure in a case where we are able to compare directly with a solution.
To this end, we used a quark--anti-quark BSE in rainbow-ladder approximation that is reviewed in Ref.~\cite{Eichmann:2016yit}.
The comparison between the extrapolated and the calculated eigenvalues is described in \ref{sec:meson}. Within errors both results
agree very well.

One complication which needs to be taken into account is that not every eigenvalue curve in the space-like domain can be extrapolated to a 
physical state with a positive real mass. We encountered (and discarded) some complex eigenvalue curves and one example for a (spurious)
tachyonic state in case of the pseudoscalar glueball. Furthermore, eigenvalue curves may cross. Hence, the hierarchy of eigenvalues at space-like momenta does not need to correspond to the hierarchy of the solutions.

The extrapolation is illustrated in \fref{fig:extrapolation_eigenvalues} where the solutions for off-shell values of $P^2=-M^2$ are shown.
We chose deliberately to print the eigenvalues over $M^2$ and not $M$, as the extrapolation is done in $P^2$.
For this plot, \eref{eq:eigenvalue_eq} was solved for various values of $\lambda\neq1$.
As explained above, the errors are calculated by averaging over 100 random subsets of the input points.
The Schlessinger extrapolation sometimes contains artifacts, e.g., very narrow poles.
These can affect the average locally which can lead to the small roughness observed in the errors.
It might be possible to refine the averaged extrapolations by applying further statistical methods or other optimization techniques as, for example, in Ref.~\cite{Binosi:2019ecz}.
For now, the obtained precision is sufficient.

\begin{table*}
	\begin{center}
		\begin{tabular}{|l||c|c|c|c|c|c|c|c|}
			\hline
			&  \multicolumn{2}{c|}{\cite{Morningstar:1999rf}} & \multicolumn{2}{c|}{\cite{Chen:2005mg}} & \multicolumn{2}{c|}{\cite{Athenodorou:2020ani}} & \multicolumn{2}{c|}{This work}\\   
			\hline
			State &  $M\, [\text{MeV}]$& $M/M_{0^{++}}$ & $M\, [\text{MeV}]$& $M/M_{0^{++}}$  &  $M\, [\text{MeV}]$& $M/M_{0^{++}}$ & $M\,[\text{MeV}]$ & $M/M_{0^{++}}$\\   
			\hline\hline
			$0^{++}$ & $1760 (50)$ & $1(0.04)$ & $1740(50)$ & $1(0.04)$ & $1651(23)$ & $1(0.02)$ & $1850 (130)$ & $1(0.1)$\\
			\hline
			$0^{^*++}$ & $2720 (180)$ & $1.54(0.11)$ & -- & -- & $2840(40)$ & $1.72(0.034)$ & $2570 (210)$ & $1.39(0.15)$\\
			\hline
			\multirow{2}{*}{$0^{^{**}++}$} & \multirow{2}{*}{--} & \multirow{2}{*}{--} & \multirow{2}{*}{--} & \multirow{2}{*}{--} & $3650(60)^\dagger$ & $2.21(0.05)^\dagger$ & \multirow{2}{*}{$3720 (160)$} & \multirow{2}{*}{$2.01(0.16)$}\\
			& & & & & $3580(150)^\dagger$ & $2.17(0.1)^\dagger$ & &\\
			\hline
			$0^{-+}$ & $2640 (40) $ & $1.5(0.05)$ & $2610(40)$ & $1.5(0.05)$ & $2600(40)$ & $1.574(0.032)$ & $2580 (180)$ & $1.39(0.14)$\\
			\hline
			$0^{^*-+}$ & $3710 (60)$ & $2.1(0.07)$ & -- & -- & $3540(80)$ & $2.14(0.06)$ & $3870 (120)$ & $2.09(0.16)$\\
			\hline
			\multirow{2}{*}{$0^{^{**}-+}$} & \multirow{2}{*}{--} & \multirow{2}{*}{--} & \multirow{2}{*}{--} & \multirow{2}{*}{--} & $4450(140)^\dagger$ & $2.7(0.09)^\dagger$ & \multirow{2}{*}{$4340 (200)$} & \multirow{2}{*}{$2.34(0.19)$}\\
			& & & & & $4540(120)^\dagger$ & $2.75(0.08)^\dagger$ & &\\
			\hline
		\end{tabular}
		\caption{Ground and excited state masses $M$ of scalar and pseudoscalar glueballs. Compared are lattice results from \cite{Morningstar:1999rf,Chen:2005mg,Athenodorou:2020ani} with the results of this work.
			For \cite{Morningstar:1999rf,Chen:2005mg}, the errors are the combined errors from statistics and the use of an anisotropic lattices.
			For \cite{Athenodorou:2020ani}, the error is statistical only.
			In our results, the error comes from the extrapolation method.
			All results use the same value for $r_0=1/(418(5)\,\text{MeV})$, see text for details.
			The related error is not included in the table.
			Masses with $^\dagger$ are conjectured to be the second excited states.}
		\label{tab:masses}
	\end{center}
\end{table*}
\begin{figure*}[tb]
	\includegraphics[width=0.49\textwidth]{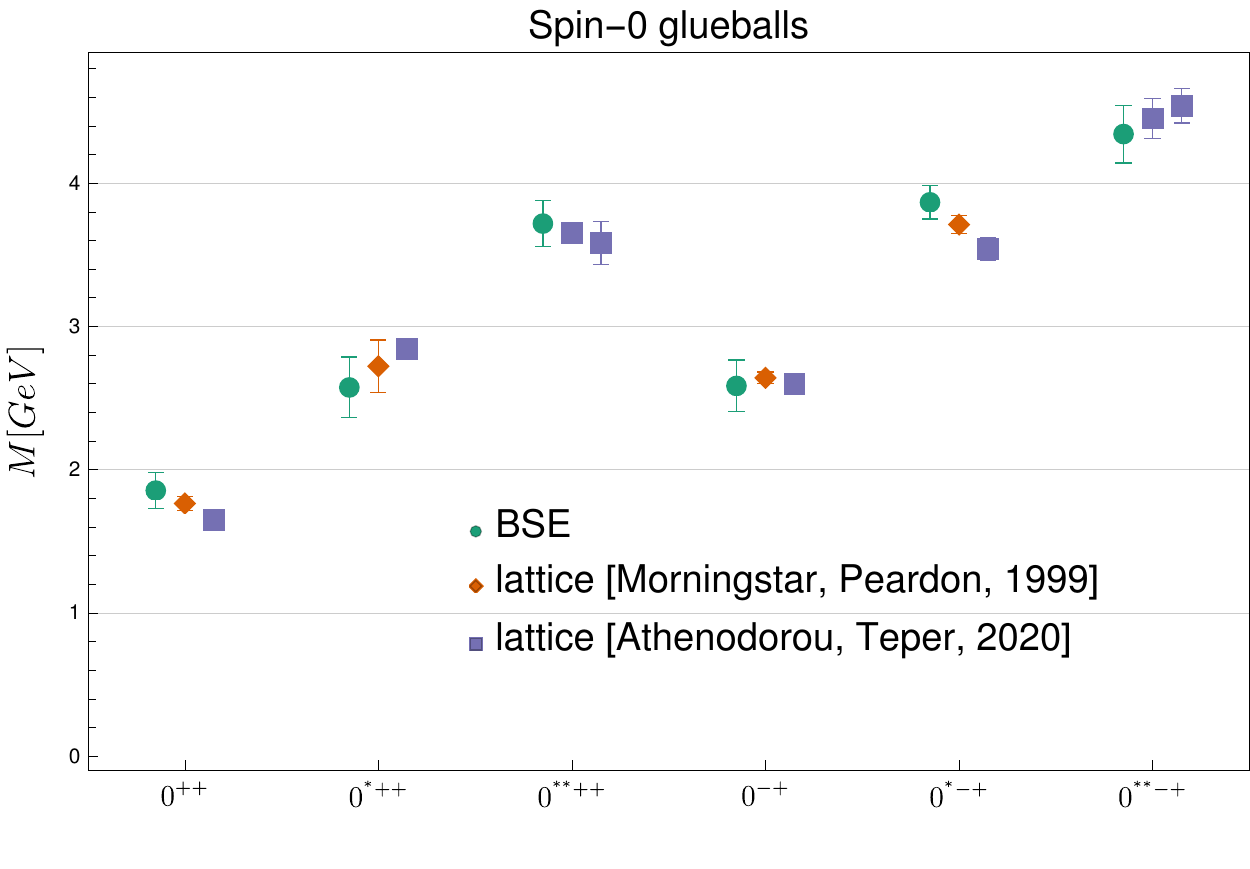}\hfill
	\includegraphics[width=0.49\textwidth]{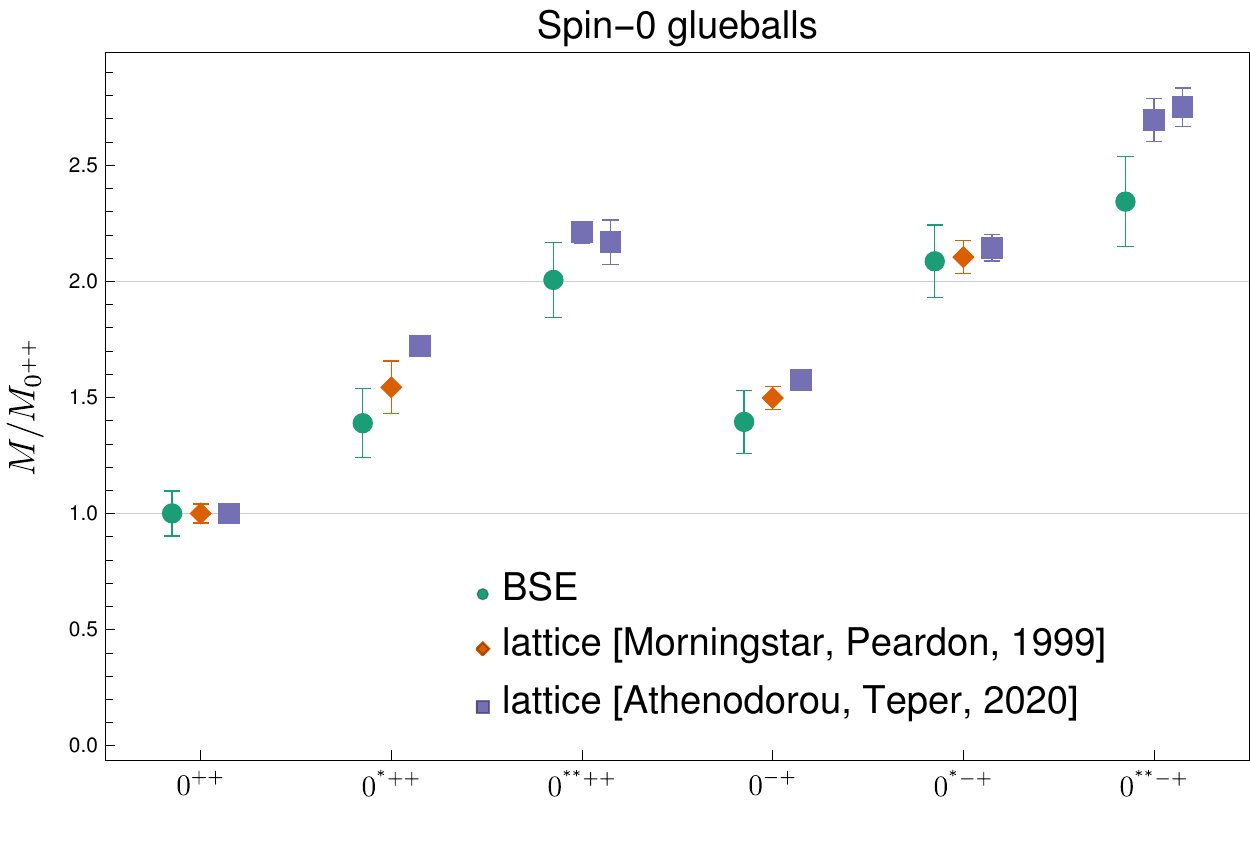}
	\caption{
		Results for scalar and pseudoscalar glueball ground states and excited states from lattice simulations \cite{Morningstar:1999rf,Athenodorou:2020ani} and this work. In the left diagram we display the glueball masses on
	    an absolute scale set by $r_0=1/418(5)\,\text{MeV})$. In the right diagram we display the spectrum relative to the 
    ground state.}
	\label{fig:spectrum}
\end{figure*}

Finally, the scalar glueball BSE leads to some additional insight with regard to the ultraviolet behavior of the Bethe-Salpeter amplitudes.
The integral in the BSE is convergent only if the amplitude falls off polynomially at large momenta. This is a standard behavior of
all Bethe-Salpeter amplitudes in the quark sector \cite{Eichmann:2016yit} and also expected for the glueball one. However, we do find eigenvalue 
curves related to states with amplitudes that run logarithmically and therefore do not satisfy this constraint\footnote{One possible reason for the appearance of such solutions is the technical implementation of the BSE problem. In practice, a cutoff for the radial integral in the BSE and discrete quadrature rules are used. Thus, the eigenproblem solved is that of a large but finite matrix which may contain additional solutions not corresponding to a solution of the original BSE and not constrained by the convergence properties stated above.}.
As a consequence, these eigenvalue curves are cut-off dependent. In the gluon DSE, a similar problem is associated with terms that break gauge covariance and lead to quadratic divergences.
A range of methods have been applied to remove these terms, see e.g. \cite{Huber:2014tva} and references therein.
In the BSE, the simplest way to deal with this problem is to identify the solutions with the wrong asymptotics as artifacts and discard them.
We adopted this strategy in this work for both scalar and pseudoscalar glueballs.
It remains to be seen whether the problem disappears entirely in even more advanced truncation schemes.

\section{Results and discussion} 
\label{sec:results}

We show our results for the ground and excited states of scalar and pseudoscalar glueball masses in \tref{tab:masses} and \fref{fig:spectrum} together with results from lattice calculations.
After the first version of this article appeared on \href{https://arxiv.org/abs/2004.00415/v1}{arxiv.org}, new lattice data became available \cite{Athenodorou:2020ani} which we also include now in the comparison.
For the sake of comparability, we adjusted all scales to the same value of the Sommer parameter $r_0$.
Originally, we set the scale of our results from the gluon propagator of Ref.~\cite{Sternbeck:2006rd} where $r_0=0.5\,\text{fm}$ was used, while for the lattice results of Refs.~\cite{Morningstar:1999rf,Chen:2005mg} it was $r_0=0.481(23)\,\text{fm}$.
In Ref.~\cite{Athenodorou:2020ani}, $r_0$ was set to $0.472(5)\,\text{fm}$ to express the obtained results in physical units.
Here, we adopt this value for the comparison of all results and rescaled the masses accordingly.
For our results this means that we multiply the masses by the factor $1.059$. 
In addition, we show the results in terms of the corresponding scalar ground state masses.

For the scalar glueball we find a ground state mass of $1.850 (130)\,\text{GeV}$.
The extrapolation is very stable in this case.
The first excited state is at $2.570 (210)\,\text{GeV}$, and we also find a candidate for the second excited state at $3.720 (160)\,\text{GeV}$.
The pseudoscalar glueball ground state is at $2.580 (180)\,\text{GeV}$.
The first two excited states are at $3.870 (120)\,\text{GeV}$ and $4.340 (200)\,\text{GeV}$.
The masses for the scalar and pseudoscalar second excited states were, to our knowledge, predicted here for the first time.

Ref.~\cite{Athenodorou:2020ani} contains lists of states for the different irreducible representations of the octahedral subgroup to which the rotational symmetry is broken on the lattice.
The spin of a glueball state is identified by searching for nearly degenerate states in corresponding representations.
The scalar and pseudoscalar glueballs only appear in the representation $A_1$.
Four states with the corresponding quantum numbers $P=1$, $C=\pm 1$ are listed.
The lower two are identified as the (pseudo)scalar glueball and its first excitation.
The other two are very close to each other.
Since their values are similar to the ones found here for the second excited states, we conjecture that one of each pair indeed is such a state.

The comparison with the lattice results of Refs.~\cite{Morningstar:1999rf,Chen:2005mg,Athenodorou:2020ani} shows that we do not only recover the same 
hierarchy but that our results also agree quantitatively very well with the lattice results even including second excited states.

\begin{figure*}[tb]
 \includegraphics[width=0.32\textwidth]{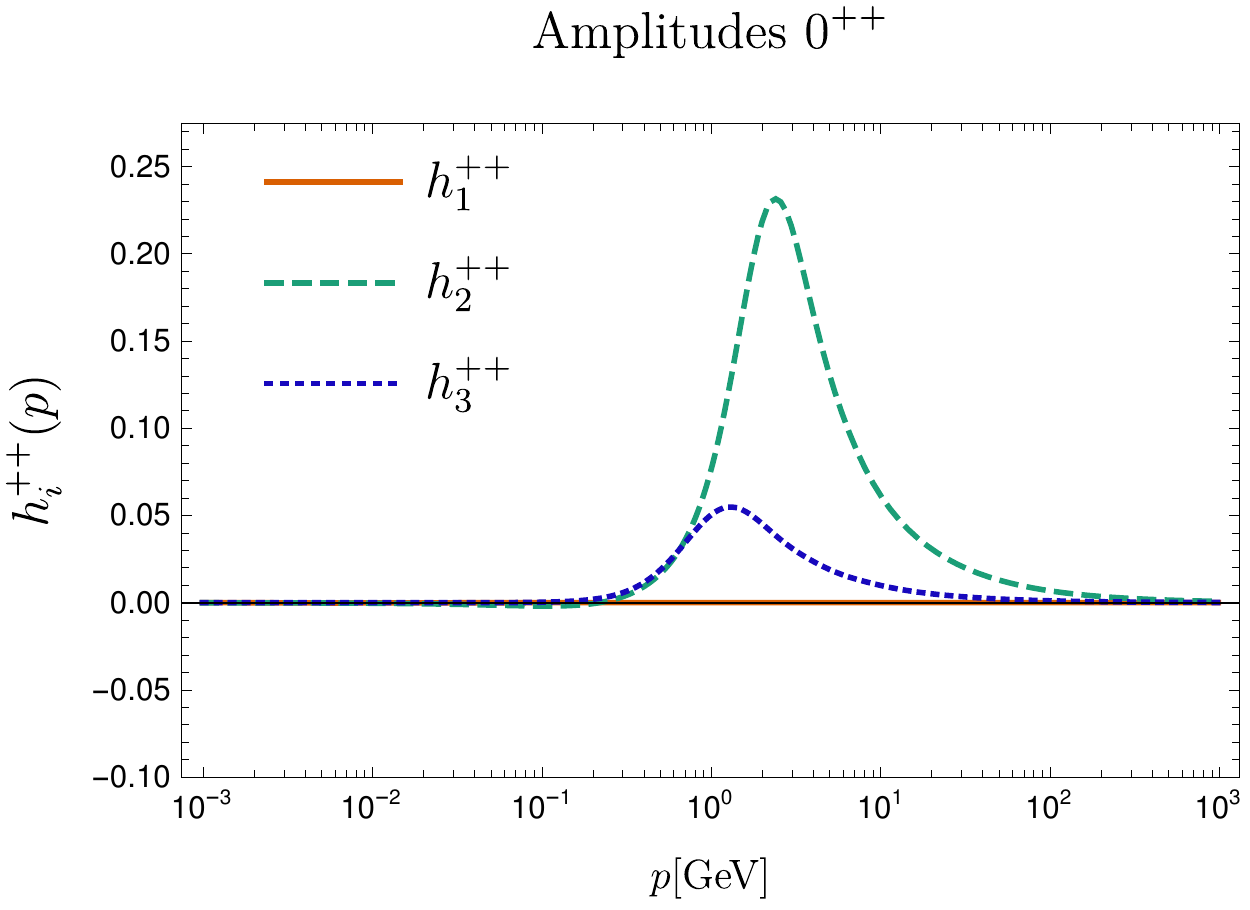}\hfill
 \includegraphics[width=0.32\textwidth]{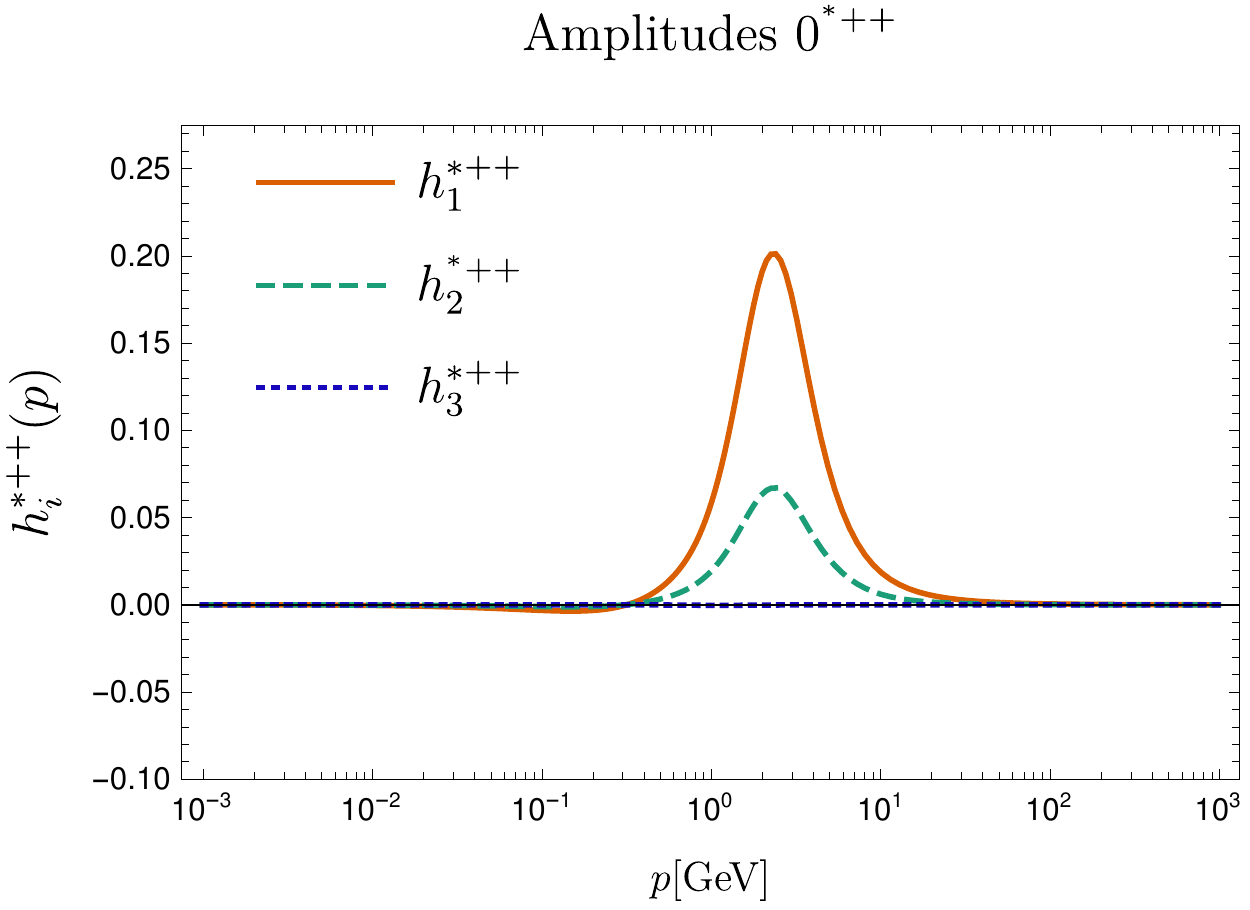}\hfill
 \includegraphics[width=0.32\textwidth]{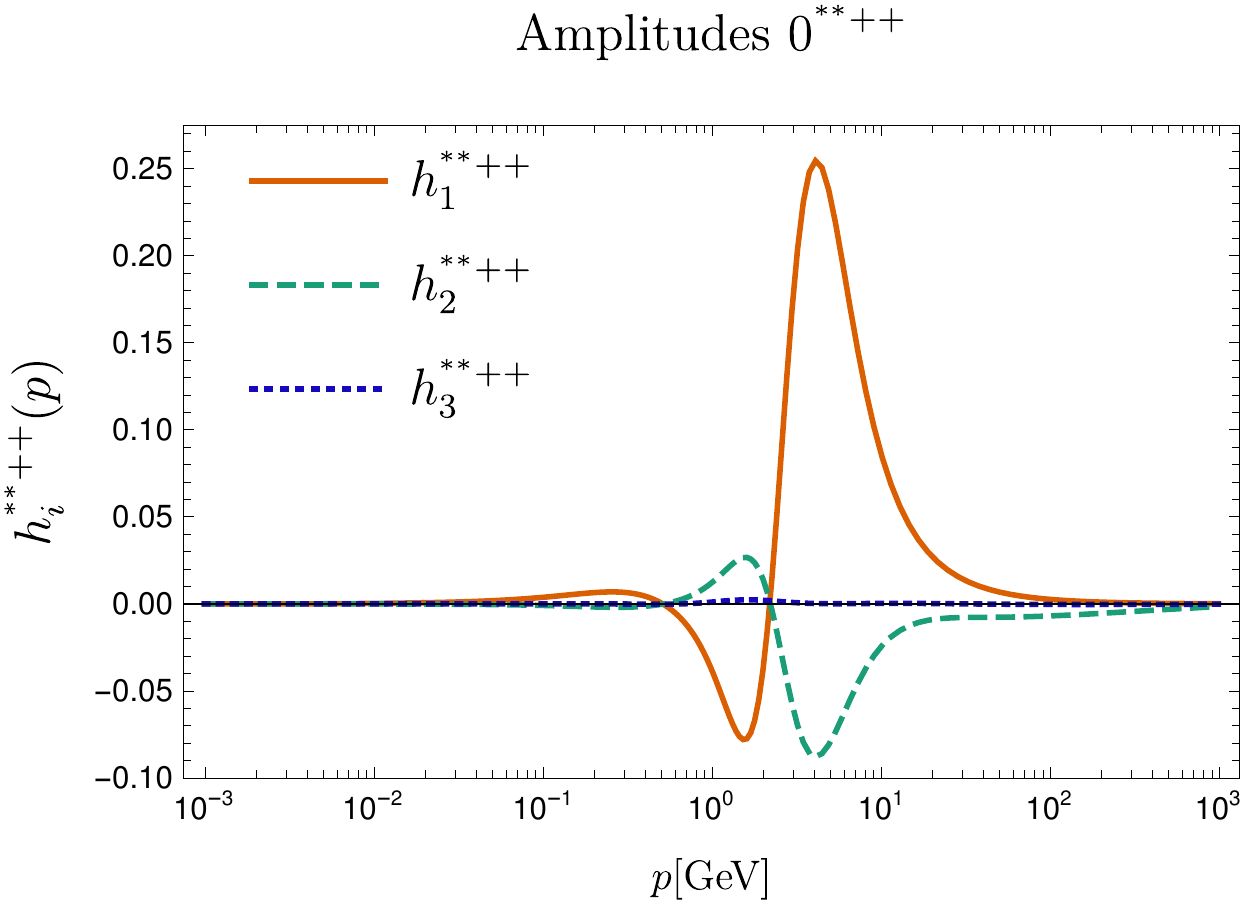}\\
 \vskip2mm
 \includegraphics[width=0.32\textwidth]{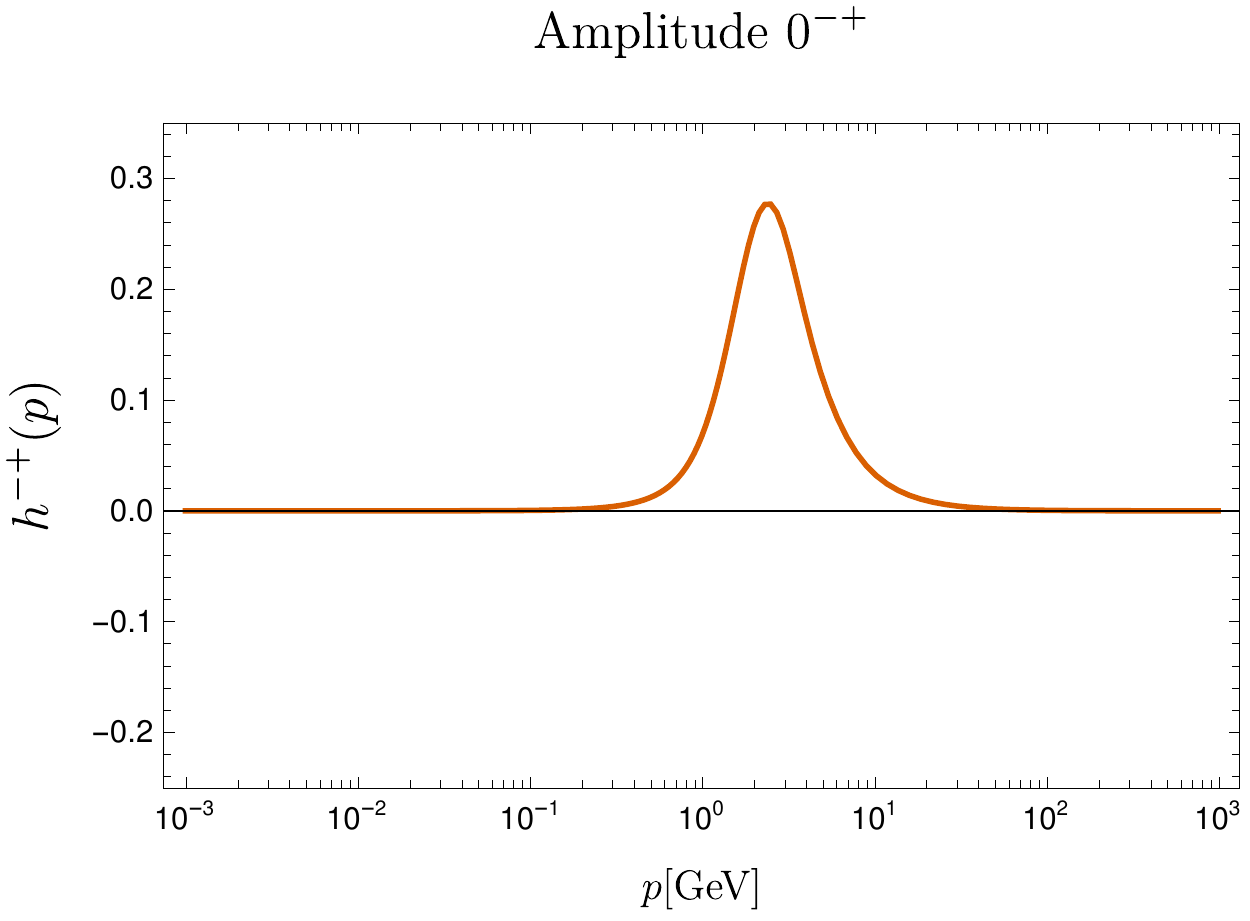}\hfill
 \includegraphics[width=0.32\textwidth]{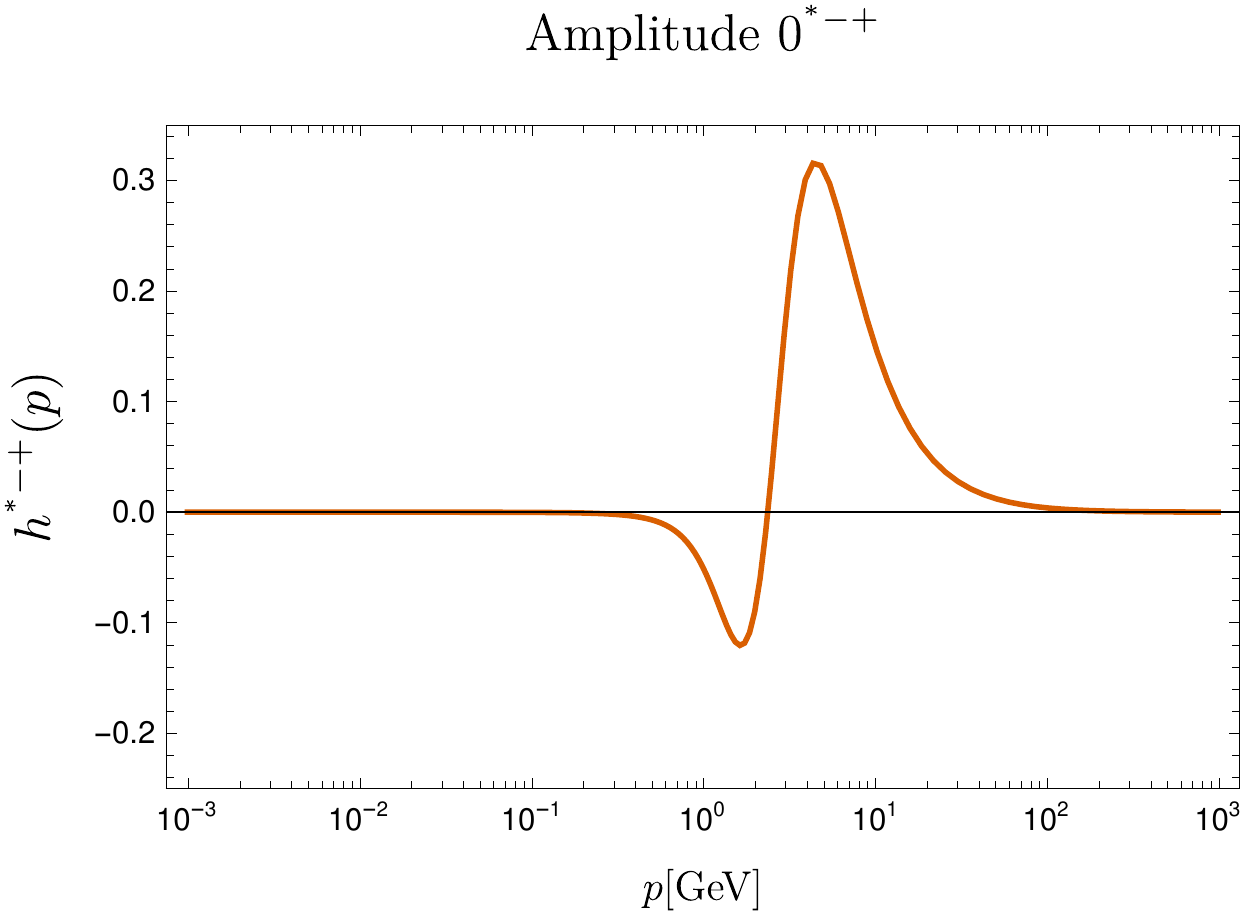}\hfill
 \includegraphics[width=0.32\textwidth]{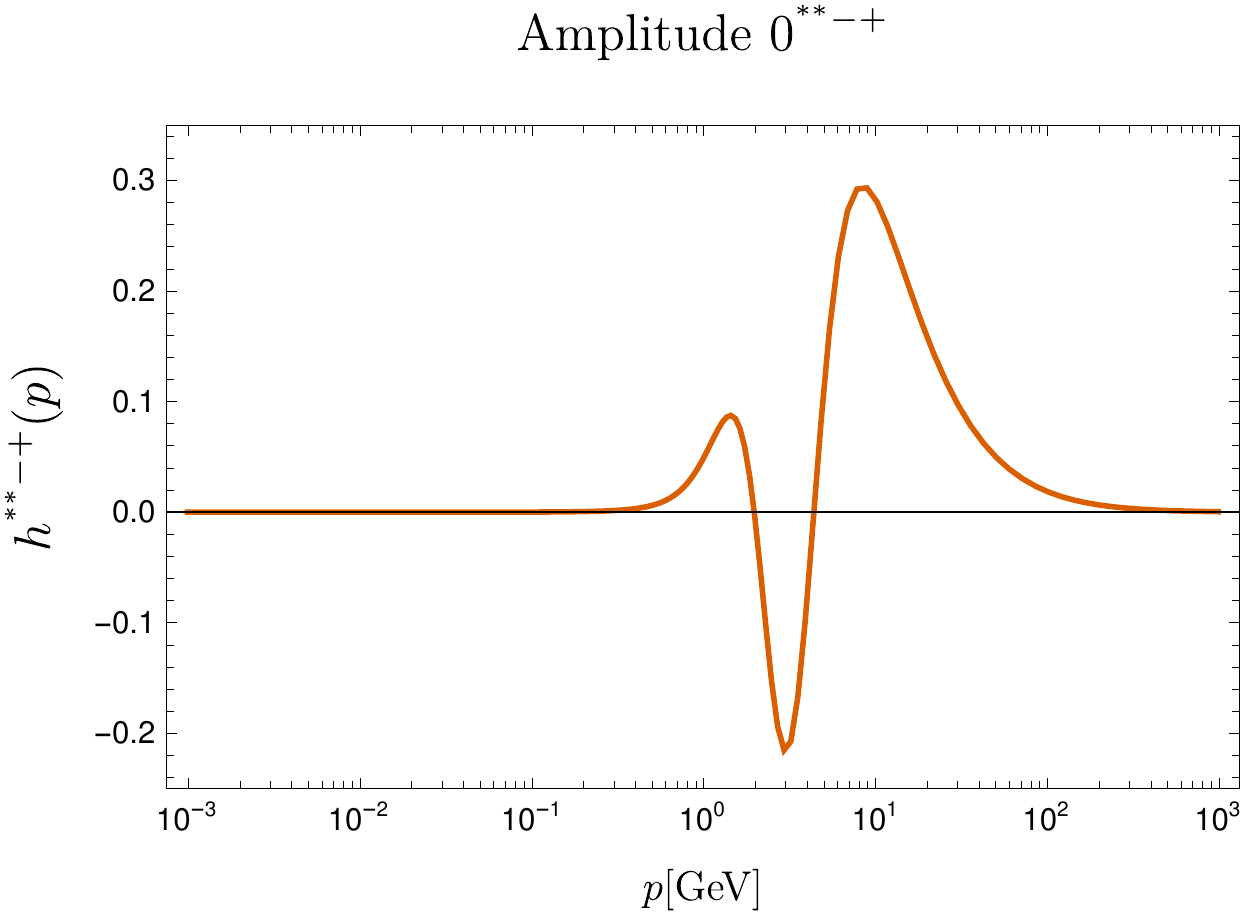}\hfill
 \caption{Leading amplitudes for the scalar (top) and pseudoscalar (bottom) ground states (left), first excited states (middle) and second excited states (right) for the lowest calculated $P^2>0$.}
\label{fig:amplitudes}
\end{figure*}

The amplitudes for the ground states and the two excited states at the lowest calculated value for $P^2>0$ are shown in \fref{fig:amplitudes}.
Although they do not correspond to the physical amplitudes, for which $P^2=-M^2$, they provide a lot of insight.
First of all, we see for the meson example described in detail in \ref{sec:meson} that the amplitudes show only a small dependence on the momentum variable $P^2$.
In that case, we can even use the results at space-like momenta to extrapolate the results to time-like momenta.
Second, off-shell amplitudes are required for calculations where the corresponding bound states are intermediate states over which one integrates.

All amplitudes in the ground states show a distinct peak around $1\,\text{GeV}$.
The first and second radial excitations contain one and two nodes, respectively. For the scalar glueball we observe an interesting interplay of the three
different components given in Eqs.~(\ref{scalar1}) and (\ref{scalar2}). Whereas the ghostball-part plays an important role in the ground state,
it is negligible for the two excited states. Thus, neglecting the ghost contribution would distort the results quantitatively (but not qualitatively) as we tested explicitly.
Furthermore, for ground and excited states different gluon amplitudes are largest.

We explicitly verified that the masses extracted for the glueballs are invariant (within error bars) of our choice of input within the one-parameter family shown in Figs. \ref{fig:gluon}-\ref{fig:3g}.
The endpoint of this family is the scaling solution for which we obtain also the same results.
This supports the view that the family corresponds to different 
gauge choices within Landau gauge \cite{Maas:2009se}. At the same time, it constitutes a nontrivial test of the reliability of the employed 
truncation, as any inconsistency could destroy the gauge independence easily. We explicitly verified that the spectrum gets distorted if 
inconsistent input (e.g. propagators from one member of the family and vertices from another) is used. 
This sensitivity on the employed input indicates that it would be challenging to replace part or all of the input by models.

In this context, it is also interesting to mention that in the process of devising the setup employed in this work, we also tried various models for propagators and/or vertices for testing purposes.
However, not only were we never able to produce results somewhat close to the present ones, also qualitative features of the eigenvalue spectrum were different.
As an example, let us mention Ref.~\cite{Sanchis-Alepuz:2015hma}, in which, using model vertices, a reasonable result for the scalar glueball was obtained, but the pseudoscalar one was above $4\,\text{GeV}$.
We thus conclude that the quality of the input is very important for a good overall picture.

\section{Summary and outlook}
\label{sec:summary}

We calculated the masses of the three lowest states for scalar and pseudoscalar glueballs in pure Yang-Mills theory from 
propagators and vertices in the Landau gauge obtained from Dyson-Schwinger equations. We gave special emphasis to the consistency between the DSE and BSE setups.
The truncation we employed is completely self-contained and does not depend on any parameter or ansatz. The scale is inherited from the
comparison with lattice results for the gluon propagator. The gauge dependent input for the propagators and vertices translates into
invariant results for the masses within a one-parameter family of nonperturbative completions of Landau gauge. Our results agree quantitatively very well 
with the lattice results of Refs.~\cite{Morningstar:1999rf,Chen:2005mg} and we add two more states to the known spectrum of pure Yang-Mills theory.
It is encouraging that these new states can be matched with subsequently published lattice results~\cite{Athenodorou:2020ani}.

To our mind, the present results constitute a considerable step forward to describe QCD bound states in a continuum approach from first principles.
There are many applications to which this setup can and should be extended. In particular, the quark sector should be included to get access to 
real-world glueballs. The calculation of higher spin glueballs or even other exotics like hybrids may be possible within the same framework as well.
At the same time, these results motivate further studies of functional equations for complex momenta to eliminate the need of the extrapolation 
of the eigenvalue curves from space-like Euclidean data.
A step in this direction was taken in Refs.~\cite{Strauss:2012dg,Fischer:2020xnb}.

\section*{Acknowledgments}

We thank Richard Williams for useful discussions.
This work was supported by the DFG (German Research Foundation) grant FI 970/11-1, by the BMBF under contract No. 05P18RGFP1, and by the FWF (Austrian Science Fund) under Contract No. P29216-N36.
The HPC Cluster at the University of Graz was used for the numerical computations.

\appendix

\section{Comparison between calculation and extrapolation at time-like momenta for a meson example}
\label{sec:meson}

We illustrate the extrapolation procedure to time-like momenta via Schlessinger's method with the example of a meson.
For the interaction we use the Maris-Tandy model \cite{Maris:1999nt} that allows also a direct calculation so that we can compare the numerically exact results with the extrapolation.
Since we are only interested in the reliability of the method, we do not consider a specific physical example but simply use a system of a quark and an anti-quark such that we obtain a meson mass of $M=2.62$ GeV (with $\eta = 1.8$ and $\Lambda = 0.72\,\text{GeV}$ in the Maris-Tandy model).
This is in the same range as the glueball masses.
Our extrapolation procedure leads to $M=2.67(4)$ GeV.

\begin{figure}[tb]
	\includegraphics[width=0.49\textwidth]{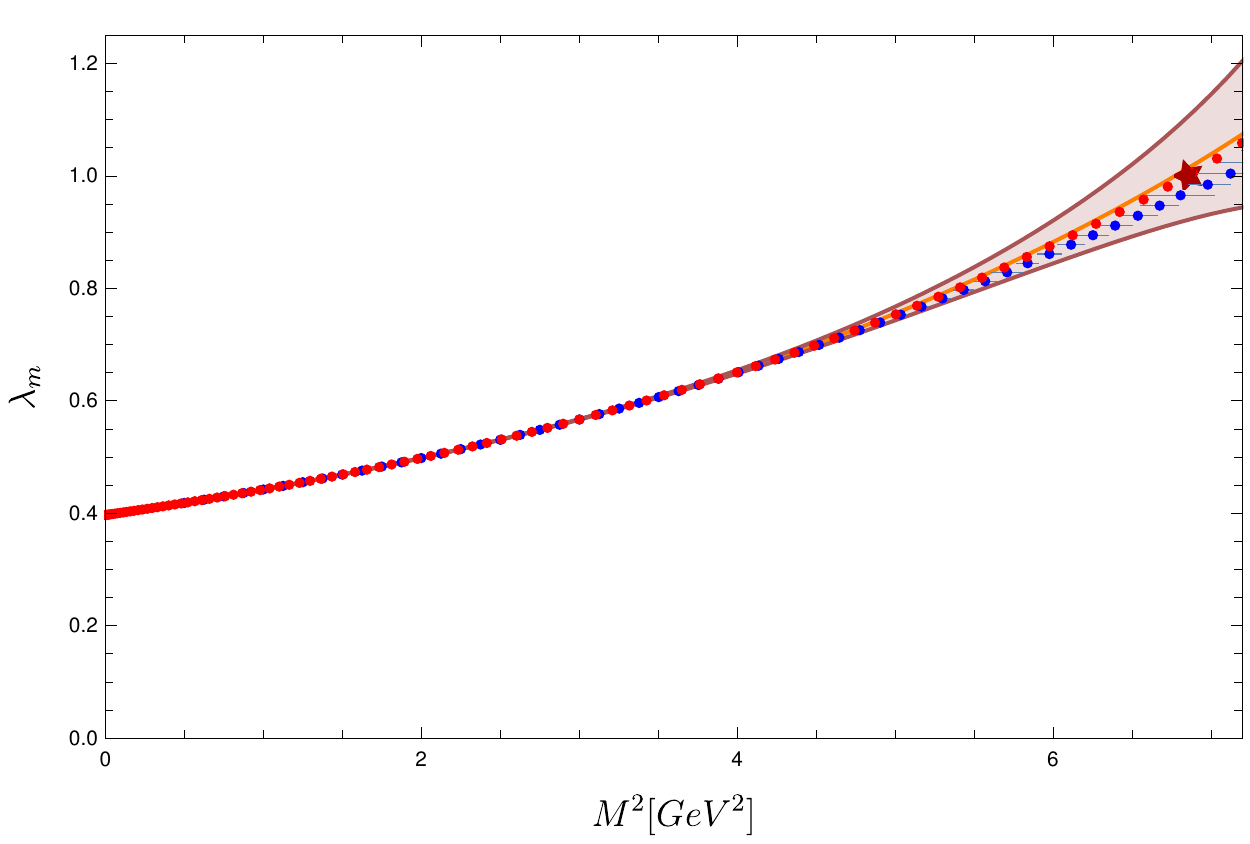}
	\caption{Extrapolation exemplified with a meson of mass $M=2.62\,\text{GeV}$.
        The red dots represent the exact solutions for the eigenvalues, the star the physical one.
		The orange line is the averaged extrapolation with errors indicated by the band.
		The horizontal error bars represent the errors for specific mass values.
		Below $2\,\text{GeV}$, the agreement is so good that the points lie on top of each other.}
	\label{fig:meson_extrap}
\end{figure}

\begin{figure}[tb]
	\includegraphics[width=0.49\textwidth]{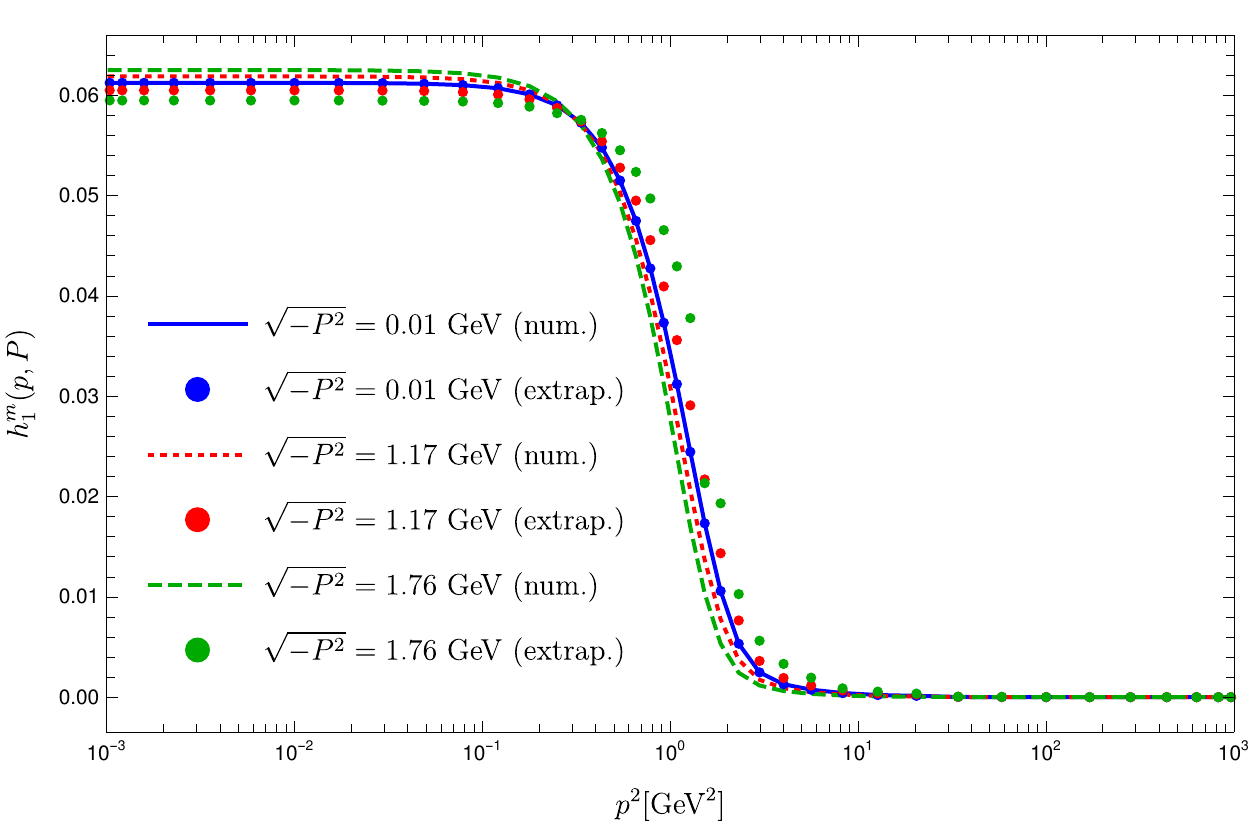}\vspace*{5mm}\\
	\includegraphics[width=0.49\textwidth]{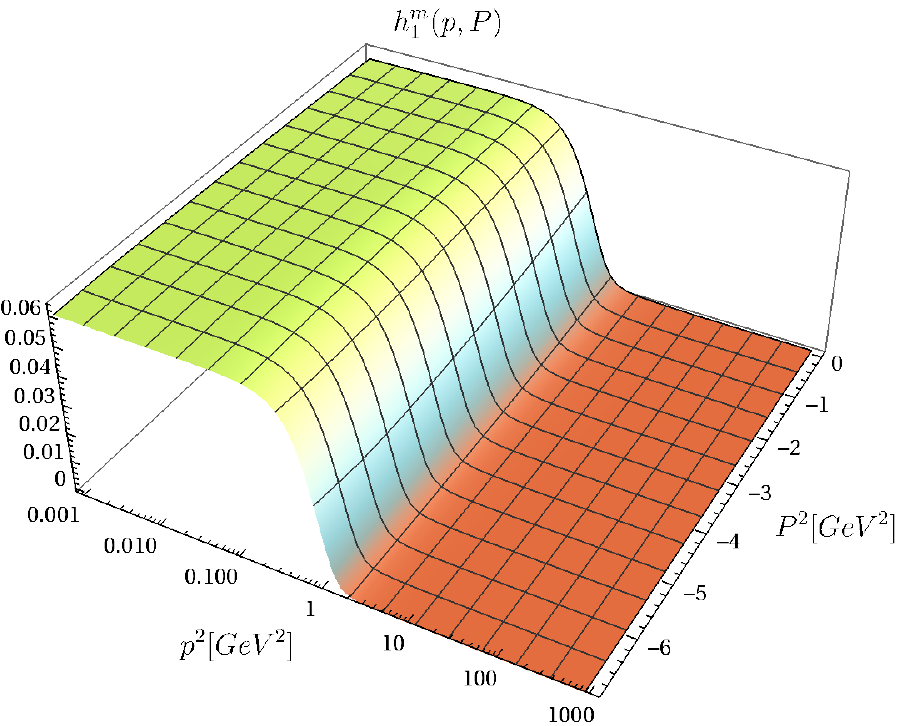}
	\caption{Top: Comparison of calculated (line) and extrapolated (dots) leading meson amplitudes at selected values of $P^2$.
	Bottom: Leading meson amplitude at time-like momenta~$P^2$.}
	\label{fig:meson_amp_extrap}
\end{figure}

We illustrate the extrapolation in \fref{fig:meson_extrap} which shows the eigenvalues calculated for real masses (red points) and the extrapolation from the calculation on the space-like side (orange line with error band).
One has to distinguish two ways of measuring errors.
The orange band in \fref{fig:meson_extrap} is calculated from averaging over extrapolations using random subsets of input points.
The mass is determined from the position where the eigenvalue curve crosses one.
To calculate an error for the mass, we average over corresponding solutions.
Since the distribution of extrapolation curves is not necessarily uniformly distributed, this error does not have to agree with the orange band.
In the plot, we illustrate this by solving for would-be masses, viz., we solve for a range of $\lambda$'s.
This gives us the blue dots with the horizontal error bars which correspond to the errors given for the glueballs results.
All indicated errors correspond to one standard deviation.

In the main text, we showed results for the glueball amplitudes at an off-shell value for $P^2$.
For the  meson example discussed here, we can explicitly illustrate that the $P^2$-dependence of the amplitude is small.
This is shown in \fref{fig:meson_amp_extrap}.
In addition, we also extrapolated the amplitude from the space-like to the time-like side and found only small quantitative differences.
Since this is sufficient for illustration purposes, we do not average over several extrapolations as for the eigenvalue curve and simply use all input points for the extrapolation.

\bibliographystyle{utphys_mod}
\bibliography{literature_glueballs}

\end{document}